\documentclass[sigconf, noacm]{acmart}
\usepackage{natbib}
\usepackage{xspace}
\usepackage{pdflscape}
\usepackage{caption}
\usepackage{subcaption}
\usepackage{svg}
\usepackage{amsmath}
\usepackage{amsfonts}
\usepackage{mathtools}
\usepackage{algorithm}
\usepackage{multirow}
\usepackage[noend]{algpseudocode}
\usepackage{hyphenat}
\usepackage{cancel}
\usepackage{breakurl}
\usepackage[english]{babel}

\usepackage{xcolor}

\algnewcommand{\LeftComment}[1]{\Statex \(\triangleright\) #1}
\algnewcommand{\IfThenElse}[3]{%
  \algorithmicif\ #1\ \algorithmicthen\ #2\ \algorithmicelse\ #3}
\algnewcommand{\IfThen}[2]{%
  \algorithmicif\ #1\ \algorithmicthen\ #2}

\newcommand{\unitCarbon}{gCO2\xspace}
\newcommand{\unitCI}{\unitCarbon/kWh\xspace}
\newcommand{\unitCIDT}{\unitCarbon/Gb\xspace}
\newcommand{\awsngcp}{AWS and Google Cloud\xspace}

\renewcommand\footnotetextcopyrightpermission[1]{} %
\setcopyright{none}
\begin{document}

\date{}

\title{The effect of the network in cutting carbon for geo-shifted workloads}

\author{Yibo Guo}
\affiliation{%
  \institution{University of California, San Diego}
  \city{La Jolla}
  \state{CA}
  \country{US}
}
\email{yig004@ucsd.edu}

\author{Amanda Tomlinson}
\affiliation{%
  \institution{University of California, San Diego}
  \city{La Jolla}
  \state{CA}
  \country{US}
}
\email{actomlin@ucsd.edu}

\author{Runlong Su}
\affiliation{%
  \institution{University of California, San Diego}
  \city{La Jolla}
  \state{CA}
  \country{US}
}
\email{r3su@ucsd.edu}

\author{George Porter}
\affiliation{%
  \institution{University of California, San Diego}
  \city{La Jolla}
  \state{CA}
  \country{US}
}
\email{gmporter@cs.ucsd.edu}

\begin{abstract}

Organizations are increasingly offloading their workloads to cloud platforms. For workloads with relaxed deadlines, this presents an opportunity to reduce the total carbon footprint of these computations by moving workloads to datacenters with access to low-carbon power. Recently published results~\cite{CRIo_eenergy23} have shown that the carbon footprint of the wide-area network (WAN) can be a significant share of the total carbon output of executing the workload itself, and so careful selection of the time and place where these computations are offloaded is critical. In this paper, we propose an approach to geographic workload migration that uses high-fidelity maps of physical Internet infrastructure to better estimate the carbon costs of WAN transfers. Our findings show that space-shifting workloads can achieve much higher carbon savings than time-shifting alone, if accurate estimates of WAN carbon costs are taken into account.

\end{abstract}

\maketitle

\section{Introduction}
\label{sec:intro}

Organizations have embraced cloud computing to a significant degree. In 2020,
spending on cloud platforms grew by 35\% to reach almost \$130 billion,
whereas spending on on-prem datacenters dropped by 6\% to under \$90 billion~\cite{srgresearch-cloudspending}.
A major reason that organizations migrate their compute and data to these
platforms is to achieve features and properties that which would be difficult or
cost-prohibitive for them to implement themselves~\cite{azure-ai-services,apple-private-cloud-compute}. Migrating work to the cloud
permits scaling up (or down) resources in response to changes in user behavior or
need. Migrating data geographically can enable replication for improved availability
and fault tolerance.  And more recently, cloud adoption can help organizations
meet climate-related goals such as limiting the carbon footprint of their
compute needs.

Cloud providers are also using the flexibility of their platforms to reduce
their own carbon footprints. Researchers have proposed~\cite{lets-wait-awhile},
and Google has publicly shown~\cite{google-datacenter-sunwind}, how to shift
workloads with relaxed deadlines forwards or backwards in time to align with
the availability of lower-carbon energy sources (such as wind and solar photovoltaic).
The net result is delivering a lower carbon footprint for the same underlying computation.
Given that low-carbon energy sources are heavily affected by environmental conditions
(the sun in the case of solar, and wind speed in the case of wind power), there is
significant potential in migrating data and workloads geographically to match
compute demands with the status of energy grids and environmental
conditions~\cite{forte-sigcomm, azure-naive-space-shifting, low-carbon-k8s-scheduler,carbon-savings-via-load-migration-caiso-pjm}.
Already, geographic migration and replication is a key enabler of reducing
wide-area latency~\cite{forte-sigcomm, SPANStore}, better managing WAN bandwidth~\cite{redesign-dc-for-renewables},
or lowering monetary cost~\cite{supercloud}.

Any optimization task requires accurate information on the
system's underlying resource utilization. Reducing the CPU requirements of a network-bound
application is unlikely to deliver results in the same way that increasing the
IO resources of a CPU-bound application will likely be unproductive. For
geographic migration of cloud workloads, accurately estimating the carbon impact of the
wide-area network (WAN) transfer is critical to reducing the overall carbon footprint of
the computation. We observe that inaccuracies or
omissions in estimating the carbon cost of the WAN lead
to migration decisions that fail to achieve carbon optimization goals, and could
even be entirely counterproductive.
For example, as we’ll show in \autoref{sec:background.motivation}, a network-heavy job can achieve $75\%$ carbon savings without considering the network carbon cost, but if we account for the network transfer cost, we may not even want to migrate at all.

In this paper, we show that previous
approaches to estimating the carbon impact of WAN transfers underestimate the
real impact by up to 2-3x due to inaccurate understandings of the underlying routes, and we describe improvements to these approaches to better capture a higher fidelity picture of the carbon costs of WAN transfers.
In particular, the contributions we claim in this paper are:
 \vskip -1em
\begin{itemize}
    \item We improve upon state-of-the-art WAN carbon transfer estimation approaches based on traceroute data (Tabaeiaghdaei et al.~\cite{CRIo_eenergy23}), as shown in \autoref{tab:route.comparison.features} and \autoref{fig:routes.methods.comparison}.
    \item We have built a dataset that maps components of WAN paths between public cloud locations to their underlying energy grids (called ISOs), and will make this publicly available for other researchers to build upon
    and to use to verify and replicate our network transfer cost results (see \autoref{sec:appendix.route-dataset}).
    \item We developed an highly optimized and efficient algorithms for carbon-aware workload scheduling between geographically distributed data centers, quantitatively analyzing the potential additional benefit of space-shifting solutions over previously published time-shifting solutions~\cite{google:carbon-aware-computing, lets-wait-awhile}.
\end{itemize}

\section{Motivations and Background}
\label{sec:background}

In this section, we first describe the high-level motivations of our work on carbon-aware workload migration, and then define the terms and definitions that we use throughout the paper; finally we show a concrete example that highlights some of the factors impacting a successful workload migration.

\subsection{Motivations}

We first describe the motivations for general carbon-aware computing in \autoref{sec:background.motivation.increasing_renewables}, and then the specific reason advocating for space-shifting in \autoref{sec:background.motivation.lower_wan_cost}.

\subsubsection{Increasing renewable deployment}
\label{sec:background.motivation.increasing_renewables}

A key trend in carbon-aware computing is to adapt to the availability of renewable energy, which varies in both time and space domain~\cite{google:carbon-aware-computing,lets-wait-awhile,carbon-savings-via-load-migration-caiso-pjm,google-datacenter-sunwind}. This means that some data centers may have access to a lot of low-carbon renewables (e.g. hydro- or nuclear-based locations) and others may have none (e.g. coal- or gas-based locations); or they may only have access to low-carbon during certain times of day or certain months of the year (e.g. solar- or wind-based locations)~\cite{google-datacenter-sunwind}.

As the amount of renewable deployment continues to grow, existing grid infrastructure already lacks the capacity to store (in battery) or move (over long-distance grid lines) these excess renewable energy, due to various practical reasons like technological limitations, regulations and monetary costs. This leads to an increasing amount of curtailment, or wasted renewable energy, especially in solar powered region like California~\cite{caiso-curtailment}.

As data centers are now accounting for an increasing and non-negligible share of global electricity usage~\cite{akcp-datacenter-energy-usage, acm_techbriefs_2021}, researchers and data center operators have looked into how data centers can help alleviate this problem by shifting their power demand in time~\cite{google-datacenter-sunwind,google:carbon-aware-computing,lets-wait-awhile} and space~\cite{azure-naive-space-shifting,carbon-savings-via-load-migration-caiso-pjm}, which is a key mechanism to achieve the 24/7 carbon-free climate goals of many major companies.

Additionally, such trend has led to closer integration of renewable power and data centers, as shown in both academic~\cite{redesign-dc-for-renewables} and industry~\cite{lancium} worlds. This further demands cost-efficient carbon-aware solutions in not only time-shifting, but more importantly space-shifting.

\subsubsection{Lower cost in WAN vs power grid}
\label{sec:background.motivation.lower_wan_cost}

A key infrastructure that we leverage in carbon-aware space-shifting is the wide area network (WAN).

Although expensive in absolute numbers and being a scare resource, WAN serves a similar role in connecting data centers as power grid serves in connecting electricity consumers, and is better in two major ways: 1) it's easier to add more capacity in WAN than in power grid due to the presence of dark fibers; and 2) it's way more energy efficient to transmit data in these fiber-optics-based networks than transmit electricity over long-distance voltage lines, which can have high transmission losses.

Part of our goal in this project to verify this high-level intuition and to accurately quantify the carbon cost of workload migration over WAN.

\subsection{Background and Carbon Modeling}

We now move onto the key metrics to quantify the carbon costs as well as our carbon cost model.

\subsubsection{Carbon Intensity of Power Grids and ISOs}
Local power grids are run by Independent System Operators (ISOs) which are responsible for generating and supplying power to the grid. This energy has a \textit{carbon intensity}, which is the grams of carbon dioxide generated per kilowatt-hour electricity (\unitCarbon/kWh)~\footnote{Carbon intensity is actually measured in gCO2\textbf{e}/kWh where the e stands for \textit{equivalent}. This value fluctuate over time due to changes in the underlying fuel mix, but is often reported hourly. This is not a direct measurement of carbon dioxide emissions, but rather a metric that normalizes different greenhouse gases by the amount of warming they contribute. We use \unitCI in this paper for simplicity.}. We take our ISO carbon intensity numbers from Electricity Maps, a publicly available data provider~\cite{emap}. We assume that datacenter power will have the same carbon intensity as the grid where it is physically located. This may not be completely accurate, as datacenter operators might have special agreements with ISOs which are not publicly available. However, this assumption is common across the community~\cite{lets-wait-awhile, google:carbon-aware-computing, CRIo_eenergy23}, and our technique does not rely on a specific data source, so the carbon intensity numbers can be updated easily.

\subsubsection{CIDT: Carbon Intensity of Data Transfer}

Similar to how we measure the carbon emissions per kilowatt-hour electricity in carbon intensity,
Tabaeiaghdaei et al.~\cite{CRIo_eenergy23} proposed the concept of Carbon Intensity of Data Transfer (CIDT), which measure the per-gigabit carbon emissions of moving data across a network path~\cite{CRIo_eenergy23}. This represents the total carbon emission per unit of data transferred across all the network devices along the path, and is usually measured in gCO2/Gb(it). We adopt the same concept in our analysis, and discuss the detailed calculation in \autoref{sec:optimization.ci_cidt}. Note that CIDT is route (location) and time dependent.

\subsubsection{Carbon Emissions of Workload Migration}
\label{sec:carbon-cost-model}

In this work, we consider both \textit{compute} and \textit{transfer} carbon costs for geographic workload migration. We focus on workload migration across large geographic regions, particularly between cloud provider's worldwide locations.

The \textit{compute carbon cost} is the emissions generated during the compute phase of a workload. This is the power consumption of computer ``billed to'' the workload multiplied by the carbon intensity of the energy where that computer is. By default, we use $5$W as an estimate for the power of a vCPU, based on recent Intel Xeon CPUs.
The \textit{transfer carbon cost} of migration we further break down into \textit{network} transfer costs and \textit{endpoint} transfer costs. Network transfer costs include all network devices enabling the data transfer, and endpoint transfer costs include the sending/receiving servers at source and destination region.
In both cases, we proportionally ``bill'' the energy cost based on the usage, which is CPU utilization for compute and bandwidth for network. We use this simple model of energy billing for our analysis, and discuss the limitation in \autoref{sec:discussion}.

\paragraph{Network Transfer Costs}

For network carbon cost, we adopted the same power modeling as~\cite{CRIo_eenergy23}, where routers consume most of the energy along the network path. They have shown that due to the high energy/bit cost of Internet core routers, higher Power Usage Efficiency (PUE) and redundancy, wide-area network carbon cost can be several orders of magnitude higher than intra-datacenter networks.

In this work, we further improve the accuracy and details of network paths, by combining their traceroute data with physical Internet infrastructure dataset. This allows us to reveal the hidden hops in traceroute measurement, and more accurately account for the energy cost of fiber-attached devices based on fiber map.

We also note that the network device powers in \cite{CRIo_eenergy23} are now more than ten years old. Since we don't know the exact equipment used in today's WAN networks, and to make a fair comparison, we used the same power numbers, but we discussed the potential improvement based on new devices today in \autoref{sec:appendix.network-power}.

\paragraph{Endpoint Transfer Costs}
We model the endpoint transfer cost as the sum of power used on each end host during the transfer. We estimate $20\%$ of a machine's power at each end as the endhost sending/receiving power, based on the networking cost model in modern data centers~\cite{barroso-datacenter-as-a-computer}, and we multiply that by the carbon intensity to get the carbon emission values.

\subsection{Motivating example}
\label{sec:background.motivation}

Now that we've described our carbon cost model, we will describe a concrete example showing the potential high impact of the network in carbon-aware workload migration.

\begin{figure}[t!]
    \begin{subfigure}{\columnwidth}
        \centering
        \includegraphics[width=\columnwidth]{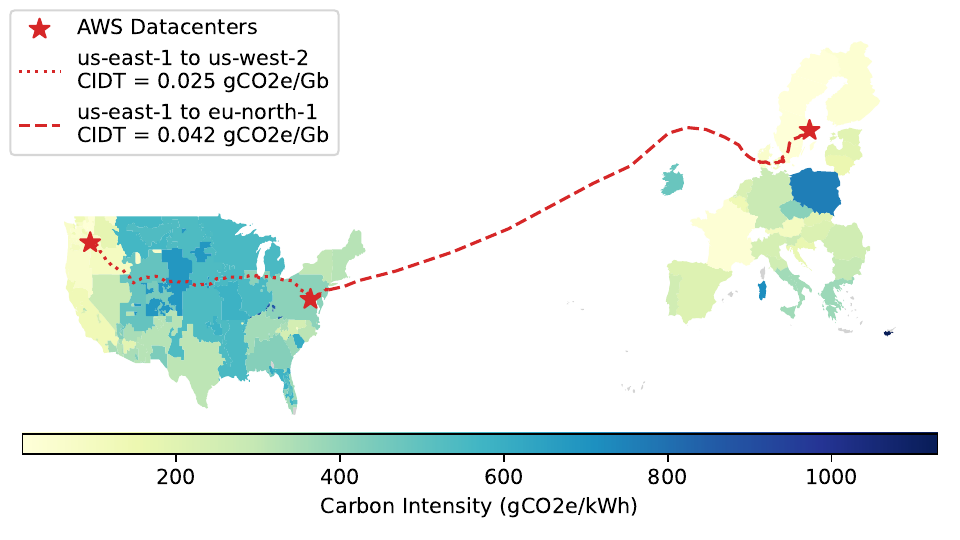}
    \end{subfigure}
\caption{
    \label{fig:motivation}
    The carbon intensity of different regions in the northern hemisphere on June 1st, 2023. The routes between AWS datacenters \texttt{us-west-2}, \texttt{us-east-1}, and \texttt{eu-north-1} are shown with their CIDTs.
    }
\end{figure}

We broadly categorize datacenter workloads into ``compute-heavy" and ``network-heavy". We first consider a network-heavy workload. Our hypothetical workload uses 10 CPU core-hours and has a total of 500 GB of input and output data. This workload is launched in Virginia, US, at the AWS \texttt{us-east-1} datacenter at noon eastern time on June 1st, 2023. 

What are the carbon emissions of this workload? Figure~\ref{fig:motivation} shows the carbon intensity (in \unitCI) of energy grid sources in the US at the time the workload is launched. AWS \texttt{us-east-1} is serviced by PJM Interconnection, which uses a combination of nuclear, coal, and natural gas. At noon on June 1st, the carbon intensity of the energy for \texttt{us-east-1} is around $400$ \unitCI. This means that a 1 hour job which consumes 1 kilowatt of power would emit 400 \unitCarbon, the equivalent of burning around $0.2$ liters of gasoline~\cite{carbon_equivalency}. Our workload which consumes 10 CPU core-hours would consume 0.05 kWh energy and emit around $20$ \unitCarbon, assuming that each core consumes $5$ Watts of power~\cite{cpu-cores-consume-5w}. 

The carbon intensity of energy in the US is not uniform, and at the same time, the carbon intensity of the AWS \texttt{us-west-2} datacenter was lower. The AWS \texttt{us-west-2} datacenter is in Washington, which is served by a combination of natural gas and hydropower. At the time our job launched, the carbon intensity of this energy is around $100$ \unitCI, which would lead to a reduction in carbon emissions of around $75\%$. If we wanted to minimize carbon emissions, it may be better to run our job in Washington, assuming there is no latency requirement.

However, this workload has input data and produces output data. To run in \texttt{us-west-2}, we would also need to transfer this data, which itself produces carbon emissions. Some of the devices on this path are served by ISOs that use relatively higher carbon energy. In this example, the CIDT of this path is $0.025$ \unitCarbon/Gb. This means that sending the data required for our job produces around 100 \unitCarbon. So even though we would save $75\%$ of the compute carbon emissions, our overall emissions would be over 5X as high, because of the high cost of transfer.

\paragraph{Compute Heavy Example:}
The previous workload is very network heavy, so we also consider a more compute-heavy workload. This workload has 100 CPU core-hours and 500 GB of input and output data. 
The AWS \texttt{eu-north-1} datacenter in Sweden has the lowest carbon intensity anywhere in the world at this time. If the network is ignored, then sending our workload to \texttt{eu-north-1} will lead to the overall lowest carbon emissions at around 10 \unitCarbon. However, to migrate our job to \texttt{eu-north-1}, we would incur around 170 \unitCarbon of migration costs, for a total of around 180 \unitCarbon. In contrast to this, the \texttt{us-west-2} datacenter would release 25 \unitCarbon for compute, but only 100 \unitCarbon for migration. Even though the carbon intensity of the energy is $\sim 2$X as high in \texttt{us-west-2}, the overall carbon emissions are around 30\% lower there.

These examples show that workload migration has the potential to reduce carbon emissions for certain types of applications, but the carbon emissions of the network cannot be ignored if we want to choose the optimal region.

\section{Related Work}
\label{sec:related}

A number of efforts have looked into selecting network paths based on carbon intensity data. Zilberman et al.~\cite{10.1145/3630614.3630618} argue for a carbon-aware network approach to considering the combined carbon of both compute as well as wide-area network transfer costs. In this work, they outline a number of metrics that are important in making carbon-aware path selections. El-Zahr et al.~\cite{10.1145/3629165} likewise consider the carbon cost of network paths, and propose a ``Carbon-aware traffic engineering algorithm'' called CATE that aims to minimize carbon emissions. As mentioned earlier in the paper, Tabaeiaghdaei et al.~\cite{CRIo_eenergy23} forecast carbon conditions along network paths and use the SCION path-aware architecture to choose paths that lower carbon emissions.

Since the availability of lower-carbon energy sources varies over time, a number of efforts have examined the ability to delay workloads with looser time requirements to times when lower carbon energy is available.  Both Meta~\cite{facebook-ML-infra,meta:carbon-explorer} and Google~\cite{google-datacenter-sunwind, google:carbon-aware-computing} explore delaying workloads, or shifting work up in time, to better match diurnal carbon cycles.
Wiesner et al.~\cite{lets-wait-awhile} focus on temporal workload shifting and identify delay-tolerant workloads that are suitable for time shifting.
Hanafy et al.~\cite{carbonscaler} exploit the elasticity in batch workloads and proposes ``carbon scaling'' using existing cloud autoscaling mechanism.

Of particular relevance to our work are approaches that migrate data and compute geographically to better match low-carbon energy sources (or alternatively, load balance requests to replicas in regions with low-carbon energy). Agarwal et al.~\cite{redesign-dc-for-renewables} propose moving computation instead of energy in response to changes in grid condition, in a group of sites that have complimentary power availability and enough WAN bandwidth.
Zheng et al.~\cite{carbon-savings-via-load-migration-caiso-pjm} show the potential carbon savings of migrating workloads between data centers in locations with different grid carbon intensities. %

From grid perspective,
Liu et al.~\cite{10.1145/2007116.2007139} explore how factoring in the carbon intensity of grid conditions on load balancing decisions can reduce the overall carbon of the composed system, and 
Lin et al.~\cite{adapting-datacenter-with-grid} further emphasize the necessity for data center operators to coordinate with the grid to achieve successful carbon savings rather than harming the grid.

At platform level,
Zhang et al.~\cite{zhang2011greenware} propose geographical load balancing algorithms that dynamically route requests to data centers with lots of renewable energy, while staying within quality of service requirements, renewable availability, datacenter capacity and WAN budget constraints.
Souza et al.~\cite{ecovisor} advocate for a new API to allow application access of carbon metrics and control of their power usage in response.
Shen et al.~\cite{10.1145/2987550.2987561} propose Supercloud, which is a live VM inter-cloud migration system that could incorporate carbon intensity as a metric for selecting where to migrate code and data.

\section{Network Carbon Costs and Routes}
\label{sec:network-route-and-carbon-data}

To calculate the carbon costs of data transfer across the wide area network, we need to know the route that data takes, along with the physical devices along that route, and the carbon intensity of the ISO powering those devices. Previous solutions~\cite{CRIo_eenergy23} have used traceroute data to infer routes, devices, and the location of those devices. In this section, we discuss this approach along with its limitations, and describe how we use additional data sources to improve our network carbon costs.

\subsection{Straw-man solution: \texttt{traceroute}}

\label{sec:route.method.logical}

Prior work~\cite{CRIo_eenergy23} calculates the carbon cost of default BGP path (for a given source and destination) by 1) running shortest path algorithm on CAIDA ITDK dataset~\cite{dataset:caida-itdk}, which is a graph of Internet routers and their connectivity, 2) translate each router IP into geolocation using the builtin IP-to-geo table, a stripped-down version of MaxMind~\cite{maxmind-geolite2}, and 3) Convert geolocation to ISOs. Total network carbon cost of the route, or CIDT, can be calculated by aggregating power usage per device multiplied by the carbon intensity at each device.

We start with the same approach to find the shortest routes across all \awsngcp regions ($19$ in AWS and $29$ in Google Cloud). We use the public IP prefixes in each \awsngcp region, and match them with the router IPs in the ITDK dataset.

However, we found that this approach, although easy to analyze and reproduce, can suffer from several limitations: 1) traceroute cannot reveal all physical hops due to the use of MPLS~\cite{caida-mpls}, especially in private links used by cloud providers~\cite{PAM2020-MultiCloud, caida-mpls-tnt}; 2) IP-to-geo datasets often lack in coverage and accuracy, resulting in low fidelity routes, and 3) without the knowledge of fiber paths, auxiliary network devices are placed like fiber amplifiers uniformly on a straight line between router hops.
Similarly, in case traceroute fails to find a path, routers are estimated based on the distance from source to destination, which can be inaccurate.

\subsubsection{Limitation $\#1$: hidden hops}

Because ITDK is based on a collection of traceroutes from public probes, it only contains routes visible on the IP layer over the public Internet. However, cloud providers often utilize private networks. Because of this and the presence of MPLS tunnels, traceroute cannot accurately reveal the physical presence of routers on a wide area route~\cite{PAM2020-MultiCloud}. From network carbon cost accounting perspective, this can significantly skew the result, as routers consume most power of a network path~\cite{CRIo_eenergy23}.

We attempted to run traceroute from within the cloud and to reveal the MPLS tunnels using \cite{caida-mpls-tnt}, but all intra-cloud traceroutes in Google Cloud show no intermediate hops between source and destination, and we revealed less than $1\%$ of the MPLS tunnels across all routes between \awsngcp regions.

\subsubsection{Limitation $\#2$: inaccurate geolocation}
\label{sec:route.method.logical.geolocation_limit}

Another issue with ITDK and traceroute is the coverage and inaccuracy of IP geolocation~\cite{geolocation-inaccurate}. This is inherent due to the nature of the data collection and is seen in all data sources.

In our analysis, we found that routes using ITDK's builtin geolocation data (\texttt{itdk.node.geo}) can often lead to unexpected locations and cause longer routes. Upon investigation and checking the source MaxMind dataset~\cite{maxmind-geolite2}, we found that these are the default values when the accuracy is low (Kansas for US, Stockholm for EU), and ITDK's geolocation data omits the accuracy radius. To get around this issue, we directly queried MaxMind and remove routes in which any router's geolocation accuracy is greater than $100$km.

The resulting set of routes still have a lot of ``noise''.
We further remove routes that do not make sense by: 1) filtering out the routes that do not start/end with the correct regions, and 2) removing routes greater than $2x$ great-circle distance.
We determine the correct start/end region using the ground truth city location of \awsngcp~\cite{cloud_gcp_regions, cloud_aws_regions}.
Finally, we break ties by choosing the most popular route that is at least $1.2x$ the minimum distance.

\subsubsection{Limitation $\#3$: no physical path}

A third disadvantage of this approach is that traceroute provides no physical path information. Not only are we ignoring the hidden routers inside private links and MPLS tunnels, we are also underestimating fiber lengths and potentially mapping fiber-attached amplifiers to the wrong ISOs. Cross-continent submarine cables also consume more power due to higher power loss, as they can only be powered from either end. This means that we are not able to accurately infer the location of auxiliary devices along the path.

\begin{table}[ht]
\begin{tabular}{c|ccc}
\hline
Information &
  traceroute &
  \begin{tabular}[c]{@{}c@{}}Inferred\\ (endpoint)\end{tabular} &
  \begin{tabular}[c]{@{}c@{}}Inferred\\ (waypoint)\end{tabular} \\ \hline
Endpoints      & x & x & x \\
\texttt{traceroute}                                               & x &   & x \\
IP-to-geo                                                         & x$^*$ &   & x \\
Infrastructure &   & x & x \\
Fiber paths                                                   &   & x & x \\
\begin{tabular}[c]{@{}l@{}}PoPs\end{tabular} &   & x & x \\
\hline
\end{tabular}
\caption{\label{tab:route.comparison.features}Comparison of different route accounting methods
\\$^*$ Note that the original traceroute data used in \cite{CRIo_eenergy23}, i.e. CAIDA ITDK~\cite{dataset:caida-itdk}, does not include IP-to-geo accuracy data.}
\end{table}

\subsection{Inferring physical routes}

From the perspective of carbon analysis, there is a pressing need for the physical location of network devices. This is because carbon metrics are based on ISOs, which map to regions of the physical world. In contrast, traceroute measures are at IP levels, and do not accurately describe the physical presence due to the aforementioned limitations.

Since we are focusing on routes over the wide area, we turn to existing Internet measurement studies to help us discover the physical location of routers on WAN links.
In particular, we use the dataset from iGDB~\cite{igdb-imc22}, a recent work on Internet route mapping using combined physical and logical measurement data, to reconstruct the routes between cloud region pairs at city-level as a graph.
We use cities of physical entities as nodes and the fiber paths among these entities as edges.
For nodes, the entities we consider include public peering locations (e.g. from PeeringDB~\cite{PeeringDB}), and optionally the Point of Presences (PoPs) of \awsngcp, to deduce this information in a reasonably accurate fashion. Our hypothesis is that even though AWS or Google Cloud can own their own fiber, the underlying physical fiber conduits are often shared, and they likely want to exchange traffic with other Internet providers at major peering cities.
 
We did, however, have to heavily re-process iGDB's dataset to connect both land and submarine fiber cables, as 1) the original fiber data for these two comes from different sources, and we need to identify land and submarine differently for power attribution, so they cannot be directly connected to build a single graph, 2) many cloud regions are not near a major peering city, so we manually ``tap'' them into nearest fibers, and 3) we also optionally add cloud provider's own Point of Presence (PoP) locations along the shortest path.

This allows us to calculate the shortest paths between any city pairs, as well as to reconstruct the physical fiber path between each hop. We refer to routes calculated using this method as \textit{inferred physical path}, or \textit{inferred path} for short.

\begin{figure}[t!]
    \centering
    \begin{subfigure}{\columnwidth}
        \includegraphics[width=\columnwidth]{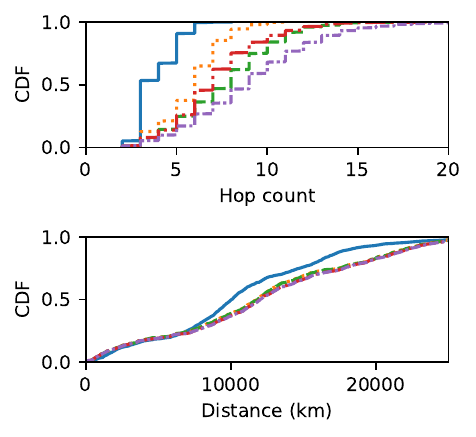}
        \caption{\label{fig:routes.methods.comparison.hop_count_and_distance} Hop count and distance}
    \end{subfigure}
    \begin{subfigure}{\columnwidth}
        \includegraphics[width=\columnwidth]{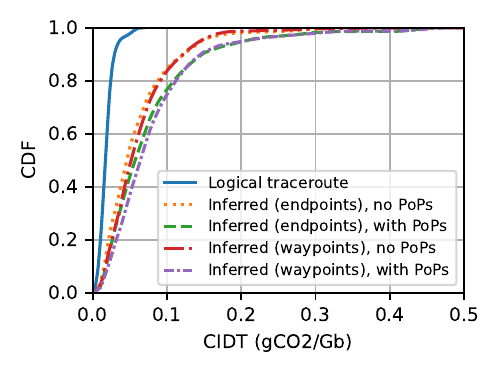}
        \caption{\label{fig:routes.methods.cidt.cdf} Network Transfer Carbon Cost in 2023}
    \end{subfigure}
    \caption{\label{fig:routes.methods.comparison} Comparison of all route accounting methods}
\end{figure}

\subsection{Combine \texttt{traceroute} and inferred paths}
\label{sec:route.comparison}

Now that we have a way to infer physical paths between two cities, we propose two strategies for using it.

One solution is that we directly infer paths based on the endpoints, i.e. source and destination region. Here, we ignore the traceroute data entirely, and purely infer the path based on the physical Internet infrastructure. We call this method \textbf{Inferred based on endpoints}.

Another solution is to use the hops in traceroute as waypoints. In this case, we filter out the inaccurate locations as described in \autoref{sec:route.method.logical.geolocation_limit}, and consider the high-confidence hops as waypoints. We can then iteratively infer the paths between each consecutive waypoints based on the approach outlined by Anderson et al.~\cite{igdb-imc22}. In this case, we consider both traceroute data and physical Internet infrastructure.
We call this method \textbf{Inferred based on waypoints}.

\subsubsection{Comparing route accounting methods}

We list the information available in each of the three route accounting methods in \autoref{tab:route.comparison.features}. Traceroute refers to the original route accounting methods described in \cite{CRIo_eenergy23} for comparison purposes.
As we can see, traceroute and inferred path based on endpoints each covers a different set of information, all of which can be useful.
For example, traceroute can indicate intermediate hops that are not on the shortest paths, possibly due to the WAN routing policy in use. On the other hand, Internet infrastructure and Point of Presence (PoP) data enable us to deduce hidden hops that traceroute cannot. Thus, we believe the combination of these two datasets, i.e. \textbf{inferred path based on waypoints is the most accurate estimate} of wide area routes for carbon accounting purpose.

We evaluated all three methods, and in the two cases of inferred route accounting methods, whether or not we include the PoPs of \awsngcp, for a total of five methods.
We performed this evaluation across all pairs of the $48$ \awsngcp regions, and plotted the distribution of hop count and distance in \autoref{fig:routes.methods.comparison.hop_count_and_distance} and the distribution of network carbon cost, or CIDT, in \autoref{fig:routes.methods.cidt.cdf}.

Note that although public cloud operators typically use private links for their intra-cloud traffic, our goal is to model scenarios where any data centers situated near these public cloud locations and thus sharing the same carbon intensity distribution, would experience in terms of WAN transfer carbon cost. Consequently, we believe our use of public Internet route data is sufficient in comparing the carbon cost and benefits of workload migration. Furthermore, private link information can be easily incorporated into our system through the Internet infrastructure database.

We can see that inferred paths generally have more hops and higher CIDT than traceroute, which is consistent with the observation of hidden hops. They also have slightly longer distances because they use accurate fiber path map instead of direct line distance.
Inferred path based on waypoints have slightly higher hop count and CIDT because we also consider the high-confidence waypoints from traceroute.
Lastly, inferred paths with PoPs have higher hop count than those without PoPs. This is because cloud providers PoPs do not show up in public peering location data. We believe adding these locations help build a more accurate view of the physical hops, as AWS traffic are highly likely to stop at an AWS Point of Presence.
\textbf{Thus for our evaluation, we choose inferred path based on waypoints with PoPs.}

Comparing the CIDT distribution with the default BGP route in Figure 4 of \cite{CRIo_eenergy23}. Tabaeiaghdaei et al. measures a median of about $0.035$ \unitCIDT, whereas we found that across all \awsngcp region pairs, the median is about $0.017$ \unitCIDT. Our result is lower but the difference is likely due to relative closer distance between the region pairs we consider vs the entire Internet analyzed in~\cite{CRIo_eenergy23}.
Comparing traceroute method against the more accurate inferred path based on waypoints method, we observe about $2-3$x difference in the median case, and much higher difference at higher percentiles.

\subsection{Carbon data availability}

As part of calculating the CIDT across all these region pairs, we discovered that the more detailed route accounting method, i.e. inferred path based on waypoints, generates a larger set of ISO due to the large amount of low-power auxiliary devices along the fiber. This creates a problem where some of the network devices are in a lesser known ISO with poor carbon data coverage.\footnote{In particular, this is because electricity map only publishes full year data for a smaller set of $130$ regions that are popular, out of the full $400$ regions around the world.}
We can discard the routes where carbon data is not available, but in the worst case, we have only half of the routes with known carbon data. Upon a deeper look, we found that in many cases, the majority of the network devices (by power) have known carbon data, which we extrapolate to the whole path.

\subsubsection{Estimating CIDT with partial carbon data}
\label{sec:carbon-data.estimation}

For each route, we first calculate the ratio of power with carbon data over the total power across all network device. If this ratio is too low, then we ignore the route as any estimation will likely incur too much error and skew our result. We considered both $50\%$ and $75\%$ cutoff threshold for this ratio.
We then compare two heuristics:
\begin{itemize}
    \item Route Average: we calculate the route average CIDT for devices with carbon data and if that's less than $100\%$, we scale this value to $100\%$.
    \item Nearest Neighbor: for each device without carbon data, we use the carbon intensity from the nearest neighbor with carbon data.
\end{itemize}

\begin{table}[]
\begin{tabular}{|lllll|}
\hline
\multicolumn{1}{|l|}{\multirow{2}{*}{Heuristic}} & \multicolumn{1}{l|}{\multirow{2}{*}{Coverage}} & \multicolumn{3}{l|}{Error Percentage (\%)}                             \\ \cline{3-5} 
\multicolumn{1}{|l|}{}                           & \multicolumn{1}{l|}{}                          & \multicolumn{1}{l|}{Mean} & \multicolumn{1}{l|}{$90\%$} & $99\%$ \\ \hline
\multicolumn{5}{|c|}{50\% Cutoff}                                                                                                                                          \\ \hline
\multicolumn{1}{|l|}{Route Average}              & \multicolumn{1}{l|}{95.6\%}                    & \multicolumn{1}{l|}{5.35} & \multicolumn{1}{l|}{19.76}     & 56.77     \\ \hline
\multicolumn{1}{|l|}{Nearest Neighbor}           & \multicolumn{1}{l|}{95.6\%}                    & \multicolumn{1}{l|}{5.22} & \multicolumn{1}{l|}{17.09}     & 58.87     \\ \hline
\multicolumn{5}{|c|}{75\% Cutoff}                                                                                                                                          \\ \hline
\multicolumn{1}{|l|}{Route Average}              & \multicolumn{1}{l|}{84.2\%}                    & \multicolumn{1}{l|}{2.56} & \multicolumn{1}{l|}{9.58}      & 28.59     \\ \hline
\multicolumn{1}{|l|}{Nearest Neighbor}           & \multicolumn{1}{l|}{84.2\%}                    & \multicolumn{1}{l|}{2.60} & \multicolumn{1}{l|}{8.86}      & 31.13     \\ \hline
\end{tabular}
\caption{\label{tab:carbon-estimation-heuristic-cutoff}Estimated network carbon cost accuracy.}
\end{table}

We use the most recent $60$ days where we have carbon data for all the ISOs to calculate the ground truth CIDT, and hide the carbon data for the ISOs which we don't have full year data to calculate the error percentage.
We summarized the coverage in $\#$ of region pairs and estimation error in \autoref{tab:carbon-estimation-heuristic-cutoff}. We found the estimation error is very similar between the two heuristics, in both $50\%$ and $75\%$ cutoff.
We chose the route average method, because our analysis found that nearest neighbor with carbon data can be hundreds to thousands of kilometers away (only $70.8\%$ coverage at $\le 500$km limit).

We do find that by limiting cutoff of known-carbon power ratio to $\ge 75\%$, we can dramatically reduce the estimation error, while still maintaining $84.2\%$ coverage (baseline without any estimation is $58.4\%$).

\section{Joint Optimization with Compute and Transfer Carbon Cost}
\label{sec:optimization}

Now, we will formally describe the mathematical optimization problem based on our carbon cost model (\autoref{sec:carbon-cost-model}).

At a high level, our goal is to find the lowest total carbon emission of a job for each region, which consists of compute, and in the case of migration, transfer carbon emission. We can then find the optimal region with lowest combined compute and transfer carbon emission.
These values depends on several factors, including the job characteristics and time-varying carbon intensity and CIDTs.

\subsection{Job characteristics}

We assume that a job has a few parameters that we know at scheduling time:
\texttt{runtime},
\texttt{core\_count},
\texttt{start\_time},
\texttt{deadline},
\texttt{input\_size},
\texttt{output\_size},
\texttt{origin}.
These parameters are often either provided by users to job schedulers, or can be profiled at the runtime, so we assume they are known.

They set the constraints of this optimization problem and determine the energy usage and the transfer time.

For our analysis, we assume a fixed network transfer bandwidth at $B = 1$Gbps, which has been demonstrated to be viable in today's commercial cloud at low cost~\cite{berkeley-skyplane}. Thus the transfer durations can be calculated as $D_{Input} = \texttt{input\_size} / B$ and $D_{Output} = \texttt{output\_size} / B$ (or $0$ for original location). Similarly, we denote $D_{Compute} = \texttt{runtime}$ as the duration of compute.

\subsection{Carbon intensity and CIDT time series}
\label{sec:optimization.ci_cidt}

We also need to know the time-varying \textit{Carbon Intensity of Electricity} (CIE) and \textit{Carbon Intensity of Data Transfer} (CIDT), in order to translate energy usage to carbon emissions, for compute and transfer respectively.

Note that both are time-dependent, and in addition, CIE is region-dependent, and CIDT is route-dependent. In our analysis, since we assume a fixed route per source/destination region pair, we can say CIDT is region-pair-dependent.

We denote $CIE_{r}(t)$ as the carbon intensity at region $r$ at time $t$, usually measured in \unitCI, and this is provided by ISOs or commercial providers like electricity map~\cite{emap}.

We denote $CIDT_{src \rightarrow dst}(t)$ as the total CIDT (in \unitCIDT) from $src$ to $dst$ across all network devices. We calculate this by aggregating the per-hop CIDT across all routers and auxiliary network devices $\mathbb{D}$, which we generated similarly as described in~\cite{CRIo_eenergy23}, but using our inferred paths:

\begin{equation*}
    CIDT_{src \rightarrow dst}(t) = \sum_{d \in \mathbb{D}}{\frac{P_{max, d}}{C_{max, d}} * CIE_{d}(t)}
\end{equation*}

Here, $P_{max, d}$ is the maximum power of a network device $d$ and $C_{max, d}$ is its maximum capacity. We use a simplified proportional energy accounting method on the network devices, i.e. each flow is ``billed'' for its share of the bandwidth used vs the total capacity of the switch.
Although other alternative options exist (e.g. marginal increase, but switches today are far from power proportional, and thus such method will still need to ``split'' the base power among all the flows),  this is outside of the scope of this work.

For submarine cables, we scale the per-distance amplifier energy cost based on its distance to the closest end of the cable, due to energy dissipation in these undersea cables, whereas we assume amplifiers for land cables are powered locally (without additional power loss).

\subsection{Optimization variables}

In time-shifting, there is one scheduling decision, that is when to start the job. Let's denote that as $t_{Compute}$.

In space-shifting, because there's also a time-varying cost of transfer based on CIDT, we also need to optimize when to start input and output transfers, which we denote as $t_{Input}$ and $t_{Output}$ respectively.

Note that input transfer, compute and output transfer must happen in order and not overlap.
This sets the constraints:

\begin{align*}
    \texttt{start\_time} &\le t_{Input} \\
    t_{Input} + D_{Input} &\le t_{Compute} \\
    t_{Compute} + D_{Compute} &\le t_{Output} \\
    t_{Output} + D_{Output} &\le \texttt{deadline}
\end{align*}

Now, given the source region $r_0 = \texttt{origin}$ and a list of alternative regions $r_i \in \mathbb{R}$, we will define the carbon emission values next.

\subsection{Compute Carbon Emission}
\label{sec:optimization.compute}

For any region $r \in \{ r_0 \} \cup \mathbb{R}$, we denote the compute carbon emission of the job (in \unitCarbon) as:

\begin{equation*}
    CE_{Compute, r} = \int_{t_{C}}^{t_{C} + D_{C}} CIE_{r}(t) * P_{C} * PUE_{DC}    
\end{equation*}

Here, compute power $P_C = \texttt{core\_count} * \texttt{watts/core}$, and $t_C$ and $D_C$ are the start time and duration of compute. We use $5$W/core as the default based on recent Xeon CPUs, but this is customizable.

\subsection{Data Transfer Carbon Emission}
\label{sec:optimization.transfer}

For the original location $r_0$, the transfer cost $CE_{Transfer, r_0}$ is zero, since we don't need to migrate the data.

For all other alternative regions $r \in \mathbb{R}$, we compute the total transfer cost as the sum of network and endpoint:

\begin{equation*}
    CE_{Transfer, r} = CE_{Network, r} + CE_{Endpoint, r}
\end{equation*}

We compute the network carbon cost $CE_{Network, r}$ using CIDT of the routes:

\begin{align*}
    CE_{Network, r} =& \int_{t_I}^{t_I + D_I} CIDT_{r_0 \rightarrow r}(t) * B \\
                     &+ \int_{t_O}^{t_O + D_O} CIDT_{r \rightarrow r_0}(t) * B
\end{align*}

Here, $t_I$ and $D_I$ are the start time and duration of input transfer, $t_O$ and $D_O$ are start time and duration of output transfer, and $B$ is the bandwidth.

We compute the endpoint carbon emission by integrating the carbon intensity with the transfer power, and we do this for both sending and receiving data, at source and destination respectively. We assume transfer power $P_{Transfer}$ to be $20\%$ of a server based on \cite{barroso-datacenter-as-a-computer}.

\begin{align*}
    CE_{Endpoint, r} =& \int_{t_I}^{t_I + D_I} (CIE_{r_0}(t) + CIE_{r}(t)) * P_{Transfer} \\
                      &+ \int_{t_O}^{t_O + D_O} (CIE_{r}(t) + CIE_{r_0}(t)) * P_{Transfer}
\end{align*}

\subsection{Optimization goal}

Given this problem formulation and constraint, the goal is to schedule the transfers and compute to get the lowest carbon emission on a per region basis.

\begin{equation*}
    CE_r = \min_{t_{I}, t_{C}, t_{O}} CE_{Compute, r} +CE_{Transfer, r}
\end{equation*}

And the best region has lowest total carbon emission:
\begin{equation*}
    \texttt{best\_region} = argmin_{r \in \{r_0\} \cup \mathbb{R}} CE_r
\end{equation*}

Note that given three optimization variables, a naive algorithm will require $O(n^3)$ time, which would be prohibitively expensive. However, we note that carbon intensity and CIDT values are only updated hourly; thus, we converted this into a discrete value search problem and developed a linear time optimization algorithm. We verified the correctness and discussed the details in \autoref{sec:appendix.optimization-algorithm}. This enables us to perform efficient analysis across the whole year at 4-hour granularity on a multi-core machine in minutes.

\section{Evaluation}
\label{sec:eval}

In this section, we evaluate the impact of geographical workload shifting using three classes of workloads that vary in their the ratio of compute to networking, which we call \textit{low}, \textit{medium}, and \textit{high} data usage workloads. Geographic migration of work is suitable for workloads with more relaxed deadlines and response time requirements. As such, we focus on batch workloads with completion deadlines in the range of minutes to hours, with a maximum completion deadline of 24 hours (similar to Radovanovic et al.~\cite{google-datacenter-sunwind}). We leave jobs with longer deadlines for future work.

\subsection{Workloads and impact on space-shifting}

In this section, we describe our selection of data center workloads and the resulting impact on the effectiveness of space-shifting solutions.

\subsubsection{Workload descriptions}

We selected representative cloud workloads that span the range of low, medium, and high data requirements. For this study, we chose (1) video resizing, (2) video processing, and (3) source code compilation, since they have relaxed completion deadline requirements and benefit from the parallelism and capability of public cloud platforms.  For each workload,
we profiled their compute requirements using Intel RAPL on a server with a Xeon Gold $6138$ CPU (with hyper-threading on) and $256$ GB of memory, running Ubuntu 20.04 LTS. We chose this configuration since it has a similar watts-per-core ratio as common virtual CPUs (vCPUs) used in \awsngcp's cloud offering. We then chose the workloads with different compute heavinesses, as shown in \autoref{tab:workloads}. We also include a workload with a low data usage but long runtime, to study the effect of runtime.

We restrict our attention to these four jobs since they span a diverse range in their data size to compute energy usage ratio, which ultimately dictates the relative cost of migration. We see in \autoref{sec:eval.job-characteristic} that the average migration overhead spans from less than $10\%$ (in the case of a compilation workload) to $100\%$ (for video resize). Beyond that point, the benefits of geographic migration diminish. We expect similar conclusions to be drawn for other practical applications when their data size to compute energy usage ratio and runtime are known.

\begin{table}[]
\begin{tabular}{lrrrrr}
\hline
Workloads            & Cores & Hours & \multicolumn{2}{c}{Input/Output} & Ratio \\ \hline
Video resize         & 4     & 2.2     & 138.0 & 7.5    & 334.6  \\
Video effect         & 8     & 2.1     & 30.0  & 3.1    & 38.8  \\
Compile (s)          & 40    & 2.2     & 24.0  & 3.9    & 6.4  \\
Compile (l)          & 40    & 11.9    & 132.0 & 21.5   & 6.4 \\ \hline
\end{tabular}
\caption{\label{tab:workloads}Workload definitions, with the last column "Ratio" representing the calculated \textit{data size to compute energy} ratio, in Gb/kWh. We use both s(hort) and l(ong) versions of code compilation to study the effect of different runtime.
}
 \vskip -1em
\end{table}

\subsubsection{Workload characteristic and impact}
\label{sec:eval.job-characteristic}

\begin{figure*}[ht!]
    \centering
    \includegraphics[width=\linewidth]{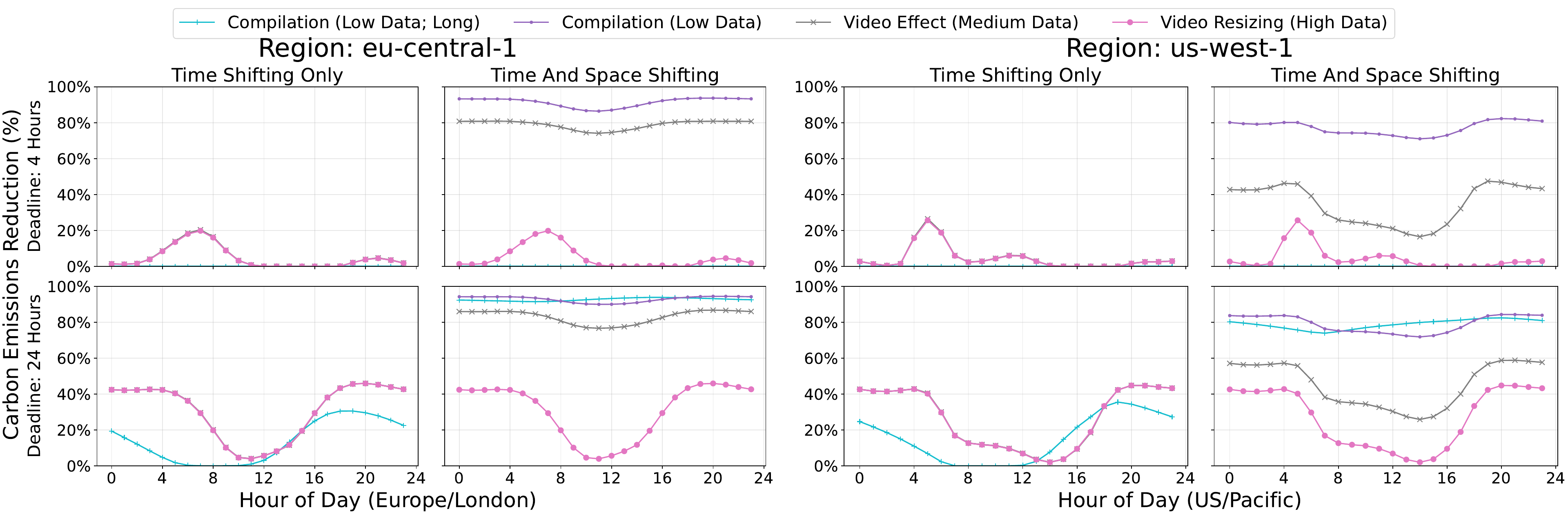}
    \caption{\label{fig:eval.time_vs_space} Comparison of time shifting vs. space shifting for two different datacenters. Time shifting is most effective for shorter jobs with long deadlines. Space and time shifting can match or beat time shifting in every scenario, with up to 90\% reduction in some cases.}
\end{figure*}

\begin{figure}[ht!]
    \centering
    \includegraphics[width=\columnwidth]{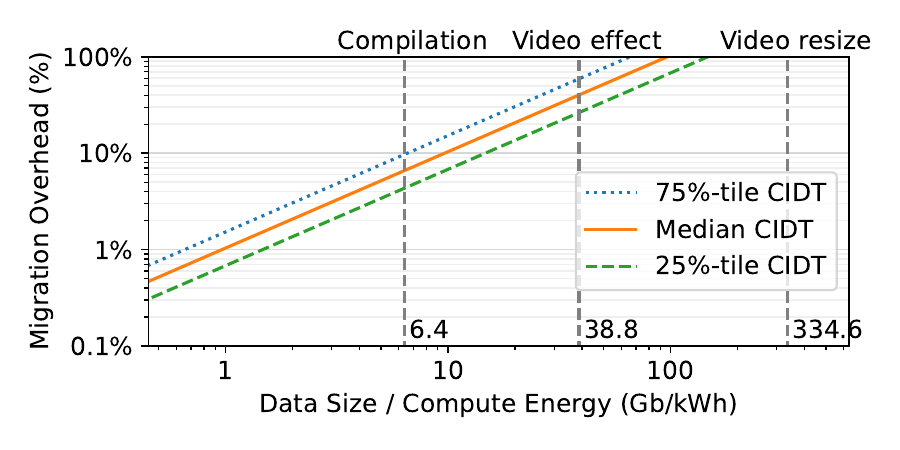}
    \caption{\label{fig:eval.jobs.migration_overhead} Migration overhead using different CIDT values
    }
\end{figure}

In this section we evaluate what kind of jobs benefit most from migration, based on their compute and data requirements. As we have shown in \autoref{sec:optimization}, the total carbon cost of a job is the sum of compute and transfer. The former is proportional to the energy usage of a job, measured in kWh, and the latter is proportional to the amount of data that needs to be transferred between regions, measured in Gb.

Thus, the net saving percentage by moving a job from $r_0$ to an alternative location $r$ is represented by:
\begin{align*}
    \text{Net saving (\%)}_r
        &= \frac{CE_{r_0} - CE_{r}}{CE_{r_0}}\\
        =& \frac{CE_{Compute, r_0} - (CE_{Compute, r} + CE_{Transfer, r})}{CE_{Compute, r_0}} \\
        =& (1 - \frac{CE_{Compute, r}}{CE_{Compute, r_0}}) - \frac{CE_{Transfer, r}}{CE_{Compute, r_0}} \\
        =& \underbrace{(1 - \frac{\int CIE_r}{\int CIE_{r_0}})}_{\text{Compute Savings (\%)}}
            -
            \underbrace{\frac{CE_{Transfer, r}}{CE_{Compute, r_0}}}_{\text{Migration Overhead (\%)}}
\end{align*}

The first part is compute savings, which is purely dependent on the carbon intensity differences, and has been analyzed in many time shifting solutions. However, migration overhead has often been ignored or simplified.

If we break this into network and endpoint, we find that:

\begin{align*}
    \text{Overhead (\%)}
        &= \frac{CE_{Transfer, r}}{CE_{Compute, r_0}}
        = \frac{CE_{Network, r} + CE_{Endpoint, r}}{CE_{Compute, r_0}} \\
        &\approx \frac{D_T * \overline{CIDT} * B + (\overline{CIE_{r}} + \overline{CIE_{r_0}}) * D_T * P_T}{D_C * P_C * \overline{CIE_{r_0}}} \\
        &= \frac{\texttt{data\_size}}{\texttt{compute\_energy}} \\
        & \hspace{2em}* \frac{(\overline{CIDT} + 1 / B * (\overline{CIE_{r}} + \overline{CIE_{r_0}}) * P_T)}{\overline{CIE_{r_0}}}
\end{align*}

$D_T = D_I + D_O$ is the total data transfer time, $\texttt{data\_size} = \texttt{input\_size} + \texttt{output\_size} = D_T * B$ is the total data size of the job, and $\overline{CIDT}$ and $\overline{CIE}$ are the average values across the integral to simplify the equation.

Note that only the first term, namely the ratio of \textit{data size} over \textit{compute energy} depends on the job, where as variables in the second term such as CIDT and CIE are job-agnostic.
Thus, from the perspective of jobs, we can see that the lower the \textit{data size / compute energy ratio}\footnote{The inverse of this concept, \textit{compute-to-data-size ratio}, and preliminary analysis of migration overhead were previously published as a poster~\cite{anonymous-poster}.}
is, the less migration overhead there is. That is to say, given a time, route and region(s), \textit{the migration overhead of a job is proportional to the ratio of data size over the compute energy}. So the question we want to answer is, when is the ratio too high or too low?

To answer this, we take the average carbon intensity of $475$ \unitCI~\cite{IEA_2019_average_ci} and median, $25\%$-tile and $75\%$-tile CIDT across all region pairs (\autoref{fig:routes.methods.cidt.cdf}), and calculate the migration overhead percentage as we vary the data size to compute energy ratio. We plot the result in \autoref{fig:eval.jobs.migration_overhead}, and overlay a few example jobs with varying ratios.
To simplify the term, we will refer to low ratio jobs as \textit{low data} jobs, as they consume relatively less data per unit of compute energy and thus incur lower relative migration carbon cost. On the other end, we call jobs with higher ratio \textit{high data} jobs; these jobs generates more carbon emissions from migration.
We can see that compilation, a low data job might incur $\le 10\%$ migration overhead, whereas a medium data job like video effect can pay around $50\%$ overhead, and a high data job like video resize might not make sense to migrate at all.

\subsection{Time shifting vs space shifting}
\label{sec:eval.time_vs_space}

Now that we have selected a set of candidate jobs and understand the relationship between compute savings and migration overhead, we can start to evaluate the effectiveness of space-shifting.

To start with, we would like to understand whether space-shifting is more or less beneficial than existing time-shifting solutions~\cite{lets-wait-awhile, google:carbon-aware-computing}.
We picked two data centers 
Figure~\ref{fig:eval.time_vs_space} shows the percent carbon emissions reduction for only time shifting, and for our approach which allows for time and space shifting. For the \texttt{us-west-1} datacenter and the \texttt{eu-central-1} datacenter, we simulated our four workloads with a 4 hour deadline and a 24 hour deadline. The workload request time was swept from 06/01/2023 to 06/29/2023 in 1 hour increments. The hourly reduction in carbon emissions were then averaged and plotted in Figure~\ref{fig:eval.time_vs_space}. Both of these regions have relatively high usage of solar energy, leading to large daily variation in carbon intensity throughout the day.

\begin{figure*}[ht!]
    \centering
    \includegraphics[width=\linewidth]{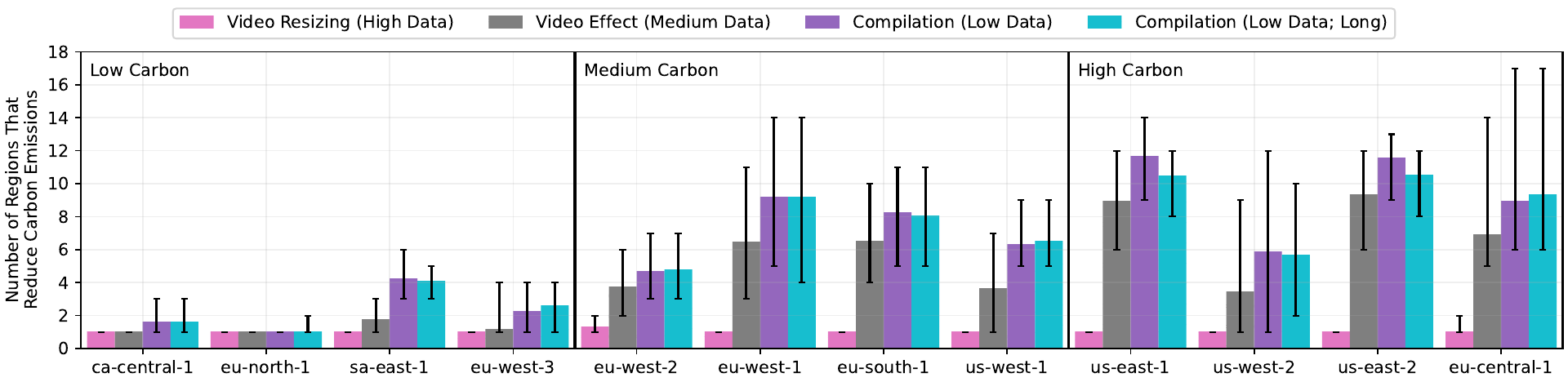}
    \caption{\label{fig:eval.region_analysis} Analysis of AWS regions that can benefit from space shifting. For each origin region, the min, average, and max number of regions that lead to an emissions reduction are shown. Regions with high average carbon intensity benefit the most from migration.}
\end{figure*}

Time shifting only can lead to carbon emissions reduction in some cases. However, this only holds for jobs that are launched outside of the low-carbon period during the middle of the day. Workloads launched at noon are already running at the optimal time and therefore do not benefit from time shifting. If we limit the deadline to 4 hours (all workloads except ``Compilation (Long)" are 2 hours), then we can only delay up to 2 hours from the request time. This reduces the effectiveness of time shifting, the only viable time to time shift is early in the morning, right before the sun comes out. Around 5-6AM in both time zones we see the largest reduction. If we increase the deadline to 24 hours after the job is launched then time shifting is more effective, but the benefit is limited to the daily variation in carbon intensity. The longer the workload is, the less effective time shifting is.

Allowing for time and space shifting leads to much better results. In the 4 hour deadline case, space shifting allows us to pick the best datacenter to run in within the deadline. The only workload that it does not help is the high data workload, and this is because this workload is not suitable for migration. Workloads that have transfer costs that eliminate the benefit of migration can still fall back on time shifting, so the performance of time and space shifting is equal to prior work in this case. Workloads that can be migrated however see up to 90\% reduced carbon emissions when compared to time shifting. In the 24 hour deadline case, even long running workloads can see reduced carbon emissions, where time shifting is not as effective.

This result shows that our method of time and space shifting can significantly reduce carbon emissions further than the state of the art time shifting only technique, especially for low data usage jobs.

\subsection{Region pair analysis}
\label{sec:eval.region-pair}

Workload migration depends both on the job-dependent migration overhead, and the time- and location-dependent CIDT and CIE. These values differ from region to region, and we expect that certain regions can benefit from migration more than others. In this section, we evaluate how many, and what types of regions can benefit. For this analysis, we roughly categorize a selection of AWS datacenters into ``low", ``medium", and ``high" carbon intensity, based on the average carbon intensity of the energy grid where the datacenter is located. We simulate our four workloads with a maximum deadline of 24 hours. For the request time, we randomly sample from the year of 2023. For our datacenters available to migrate to, we choose all datacenters in AWS. We then count the number of datacenters which reduce or keep same the carbon emissions (carbon savings of $\ge$0\%). When migrating, this number is the number of choices we have that will lead to a carbon savings of $\ge$0\%.

We show the average number of datacenters which have $\ge$0\% carbon emissions reduction for each of the datacenters in this selection in Figure~\ref{fig:eval.region_analysis}. The bars represent the minimum and maximum count of $\ge$0\% datacenters in the sample.

We can see in Figure~\ref{fig:eval.region_analysis} that our high data workload almost never benefits from migration. Only one datacenter (the datacenter it is launched in) will lead to $\ge$0\% carbon emission reduction. Across all workloads, for low-carbon regions, there are fewer choices for migration. Often times the original region is the best region for the workload. For the medium- and high-carbon regions, there are more choices. Some notable ``medium-carbon" regions are the \texttt{eu-west-1} and \texttt{eu-south-1} regions. Because they are surrounded by low-carbon regions, they still have a large number of choices for datacenters, even though they are not high-carbon. For high-carbon regions, \texttt{us-west-2} has much lower minimum values than other regions. This region has an extreme yearly carbon intensity variation, with summer months almost completely comprised of hydropower, and other months using mostly gas. So most of the year, jobs will benefit from migration, however some months it will be best to run most workloads in the original region.

This shows that not all regions benefit from workload migration equally -- it depends on both the workload and the carbon intensity of the origin datacenter. In general, high-carbon datacenters benefit more from migration.

\subsection{Load Balancing Across Regions}

\begin{figure}[h]
    \centering
    \includegraphics[width=\linewidth]{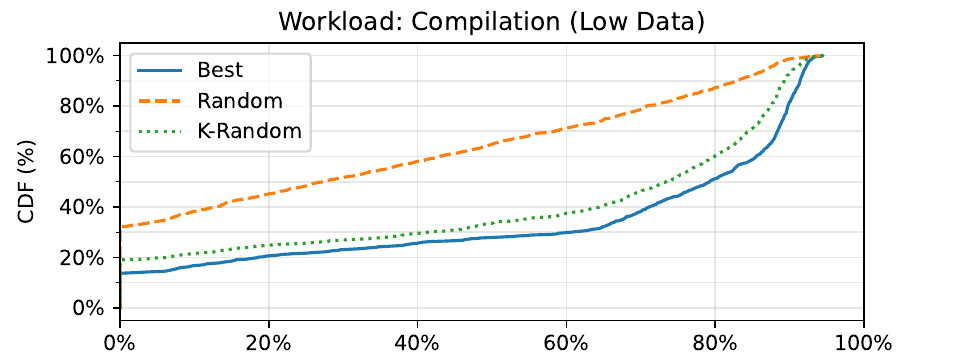}
    \caption{\label{fig:eval.load_balancing} Evaluation of some simple load balancing policies. Picking randomly from all eligible regions still leads to some reductions in emissions, however a K-Random ($K=3$) approach is almost optimal.}
\end{figure}

If we always send workloads to the datacenter with the lowest carbon intensity, we risk overloading that datacenter's capacity. Because of this, we also evaluate some simple load balancing policies. In the previous Section~\ref{sec:eval.region-pair}, we show that for each origin region there will be a set of other regions which will give $\ge$0\% carbon emissions reduction. In Figure~\ref{fig:eval.load_balancing} we evaluate three simple load balancing policies. 

The first strategy: ``Best'' is when we always choose the datacenter with the lowest carbon emissions prediction. This strategy leads to the greatest overall carbon emission reduction, however it may not be possible in practice. Since datacenters only have a certain amount of headroom, the lowest carbon datacenter may get more migration requests than it can handle. If each origin region spreads out its migration requests, datacenter headroom is less likely to run out.

The next strategy: ``Random" is randomly choosing from the set of regions that have a predicted carbon emission reduction. This spreads the migration requests across a larger pool of candidate regions. Unfortunately, not all regions may have a high carbon emission reduction. In the median case, the carbon emission reduction is only around $35\%$, compared to the best case median reduction of $80\%$.

The third strategy: ``K-Random'' is a compromise by choosing a region from the top $N$ candidate regions. This is shown for $N=3$ in Figure~\ref{fig:eval.load_balancing}. In the median case for this strategy, the emission reduction is not as good as the best case, but only by about 5-10\%. Future work can look at other load balancing policies to ensure than datacenters can always maintain a reasonable headroom.

\subsubsection{Impact at scale}

This result shows that by load balancing across several data centers with low-carbon energy, we can still achieve close to $90\%$ the best case carbon reduction, which are around $70-90\%$ for low-data usage jobs and $30-80\%$ for medium data usage jobs (\autoref{sec:eval.time_vs_space}).

If we look at existing capacity study shown in \cite{carbon-savings-via-load-migration-caiso-pjm}, with up to $30\%$ space capacity in data centers, we can achieve up to $25\%$ carbon savings for low-data usage jobs and up to $20\%$ for medium-data usage jobs. Additional data center capacity near renewable sites and/or techniques like turbo-boosting can further increase the carbon saving potentials, although their additional cost (e.g. embodied carbon footprint) may require further study.

\subsection{Time analysis}
\label{sec:eval.time}

Many datacenters are powered by renewable energy, and many datacenter operators are planning to use more renewable energy to lower overall emissions. However, renewable energy generation is highly time dependent. Solar only works when the sun is shining and wind only works when the wind is blowing. In this section we example what daily and seasonal trends may impact workload migration.

\subsubsection{Time of day / Timezone}
One downside of solar is that it cannot generate electricity during the night. However, when it's night in California, the sun is shining on the other side of the world. In this section we look at the potential to shift jobs across timezones to take advantage of where the sun is currently shining.

\begin{figure}[t!]
    \centering
    \begin{subfigure}{\columnwidth}
        \includegraphics[width=\columnwidth]{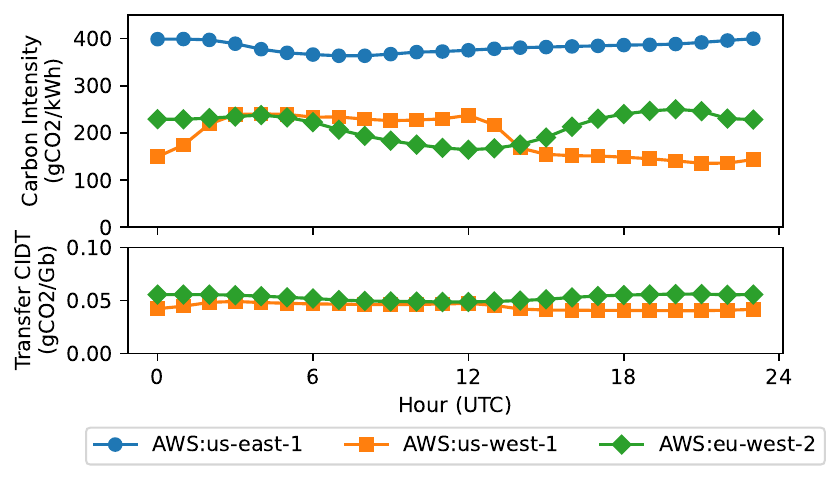}
        \caption{\label{fig:eval.timezone.carbon_intensity}Carbon intensity and CIDT time series}
    \end{subfigure}
    \begin{subfigure}{\columnwidth}
        \includegraphics[width=\columnwidth]{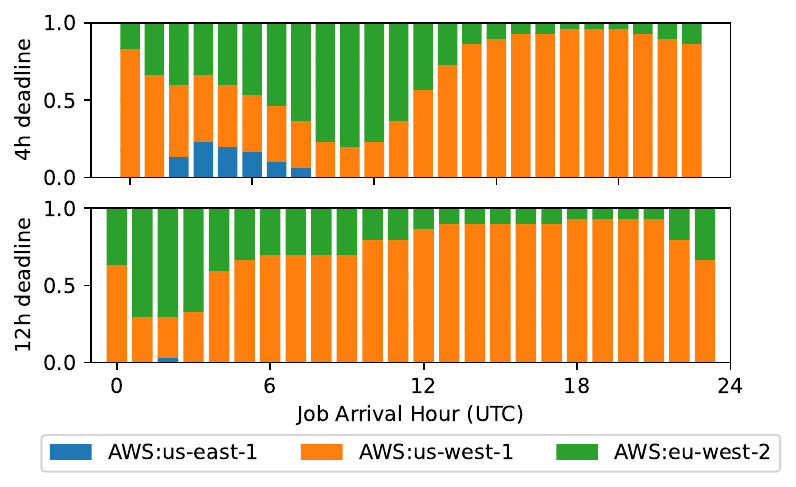}
        \caption{\label{fig:eval.timezone.optimal_region}Optimal region distribution of medium data job (video effect)}
    \end{subfigure}
    \caption{\label{fig:eval.timezone} The optimal region distribution between datacenters in different timezones. Migrating across timezones allows the utilization of solar energy almost 24 hours a day. }
\end{figure}

We consider an AWS source location in a high carbon region: \texttt{us-east-1} in Virginia, and two solar-heavy regions with daily bathtub curve, AWS \texttt{us-west-1} in California, and AWS \texttt{eu-west-2} in Great Britain. Both regions experience daily variation due to solar being available only during daytime, but are offsetted in time because of the time zone difference. The carbon intensities of both datacenters are shown at the top of \autoref{fig:eval.timezone.carbon_intensity}. The \texttt{eu-west-2} datacenter has the lowest carbon intensity around 12h UTC (12h local time), whereas \texttt{us-west-1} has the lowest carbon intensity offset by about 10 hours, around 21h UTC (13h local time). We also include CIDTs of the paths from the source location to both of these locations at the bottom of \autoref{fig:eval.timezone.carbon_intensity}. The CIDTs of both paths are comparable, since the higher energy cost of submarine cable and the shorter length of the path from \texttt{us-east-1} to \texttt{eu-west-2} cancel out each other; the curve is mostly flat due to a large number of non-renewable-based regions that the paths traverse through.

We simulate a medium data-usage job launched in the \texttt{us-east-1} datacenter. The job request time is swept across the June 2023 in 1 hour increments. We also set the maximum deadline to 4 hours and 12 hours. For each hour of day, we aggregate the optimal datacenter region over the month and show the distribution in \autoref{fig:eval.timezone}. As we can see, the optimal datacenter region changes based on the hour of the day. When carbon intensity is lowest in \texttt{us-west-1}, that datacenter is favored, however when it is night in California, \texttt{eu-west-2} is favored.
Consequently, given a deadline of h hours, jobs that arrives around (h - 2) hours before 12h UTC favor \texttt{eu-west-2}, whereas jobs that arrives (h - 2) hours before 21h UTC favor \texttt{us-west-1}.

This show that we can take advantage of this timezone difference to reduce carbon emissions acroos almost all 24 hours of a day, which is not always possible with time shifting only solutions. The geographically distributed nature of datacenters allows us to place workloads according to where renewable energy is found at any given time.

\subsubsection{Seasonal}

Another important time-varying characteristic in renewable energy is the season.
We know from \cite{lets-wait-awhile} that there are seasonal variations in the carbon intensity of solar-heavy regions. In this section we reexamine how these seasonal changes impact space shifting.

Here, we consider a Google Cloud source location in a high carbon region, \texttt{us-east4} in Virginia, and consider two alternative regions with different seasonal pattern: \texttt{us-west2} in California, and \texttt{southamerica-west1} in Chile.
We sampled the job request time at every hour across all days in 2023, and considered both a short deadline of 4 hours and a long deadline of 12 hours, for our medium data workload.

\begin{figure}[t!]
    \centering
    \begin{subfigure}{\columnwidth}
        \includegraphics[width=\columnwidth]{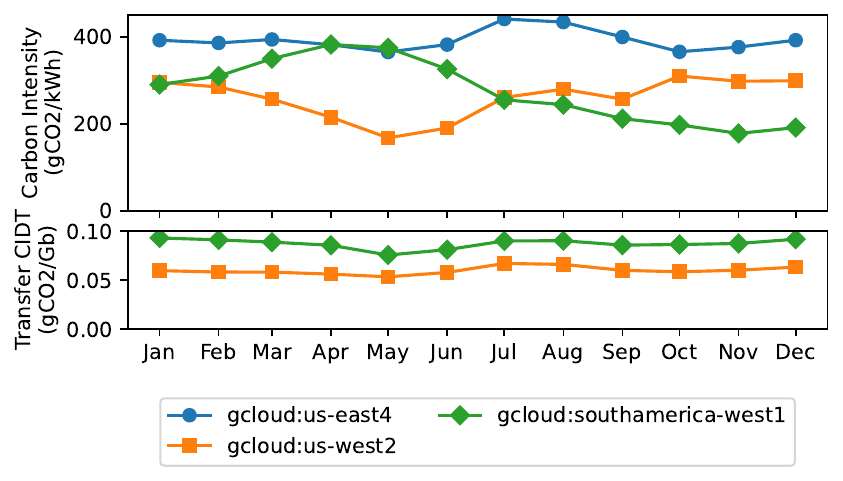}
        \caption{\label{fig:eval.seasonal.carbon_intensity}Carbon intensity and CIDT time series}
    \end{subfigure}
    \begin{subfigure}{\columnwidth}
        \includegraphics[width=\columnwidth]{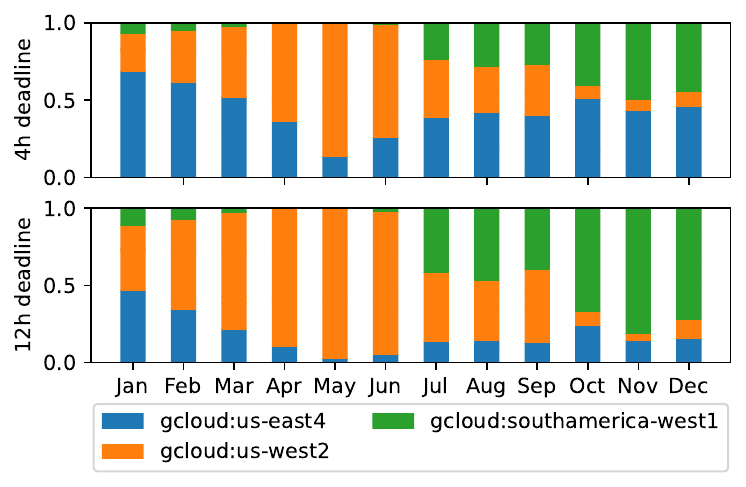}
        \caption{\label{fig:eval.seasonal.optimal_region}Optimal region distribution of medium data job (video effect)}
    \end{subfigure}
    \caption{\label{fig:eval.seasonal} The optimal region distribution across seasonal variations. In the southern hemisphere, solar energy is more productive in the winter months, the opposite of the northern hemisphere. The optimal region changes based on this effect. }
\end{figure}

We plotted in \autoref{fig:eval.seasonal.carbon_intensity} the monthly average carbon intensity in all three regions, and the monthly average CIDT from source to both alternative regions.
In this case, California has lowest average carbon intensity around May, and Chile has the lowest average carbon intensity around November. The CIDT of Virginia to Chile is generally higher than that of Virginia to California, due to the physical longer distance and use of submarine cables, but again the variations are small due to the large number of non-renewable-based regions that the paths traverse through.

In the bottom plot \autoref{fig:eval.seasonal.optimal_region}, we plotted the distribution of the optimal region across all months, for our medium data usage job (video effect). We can see that out of the two alternative regions, generally the optimal location aligns with the low carbon intensity months. Although in many cases, due to the non-trivial amount of data that needs to be moved, this job might not benefit from migration, and thus stays in Virginia. Comparing the two different deadlines, we observe that workload migration can benefit from longer deadline, since both regions are mainly powered by solar energy and thus exhibits daily shift in their carbon intensity. Thus, a longer deadline means that the job is more likely to run at the low carbon period of the day in these regions.

\section{Limitations and future work}
\label{sec:discussion}

\paragraph{Limitations:} Due to limitations in the transmission capabilities within an ISO, energy cannot necessarily freely flow throughout the ISO region, and there can be localized ``congestion'' based on transmission line limitations. In this work, we ignore these localized regions of stranded energy and instead rely on the reported carbon intensity at the ISO level. In general, better energy accounting within the datacenter would further help to select datacenter locations with greater potential for carbon savings. In our work, we did not consider the effect of network bandwidth limitations for wide-area migrations, nor did we take into account estimates of spare capacity within datacenters to ensure that workloads can ``fit'' into target datacenter locations.

\paragraph{Future work:}
We can envision an approach where application state is georeplicated to ensure that the needed data is available in regions with complimentary low-carbon energy availability. As an example, repliacs placed sufficiently far apart that solar power is available at at least one replica location. Further, we can imagine using our dataset to better inform hardware placement by putting GPUs, TPUs, etc in regions which are likely to have jobs migrated to them. Finally, we see opportunity to further develop carbon-aware load balancing approaches that select datacenters with sufficient resources to accept migrations from high-carbon regions.

\section{Conclusion}

In this paper, we show the benefits of geographically space shifting to reduce the overall carbon impact of workloads with more relaxed completion time requirements. Similar to previous studies, we have shown that the carbon cost of WAN transfers is significant, and needs to be taken into account when choosing locations for geographic load shifting. We outlined an approach and methodology for increasing the fidelity of WAN path carbon footprint estimation, and have shown that with such an estimate it is possible to reduce the carbon footprint of many workloads significantly.

\bibliographystyle{plainurl}
\bibliography{sigproc}

\appendix

\clearpage

\section{Energy Intensity of Network equipment: 2010-era vs now}
\label{sec:appendix.network-power}

We compare the 10-year old numbers from Table $2$ of \cite{CRIo_eenergy23}, i.e.~\cite{van2012power}, and newer numbers, where the main difference are the improvement of core routers from $10$W/Gbps to $1-2$W/Gbps~\cite{power-of-new-core-routers} and that of transceivers from $1.5$W/Gbps to $0.09$W/Gbps~\cite{power-of-new-transceivers}.

\begin{figure}[h]
    \centering
    \begin{subfigure}{0.9\columnwidth}
        \includegraphics[width=\columnwidth]{figures/route_methods/cidt.2023.network.avg.cdf.all_clouds.all_clouds.all_route_methods.pdf}
        \caption{\label{fig:routes.appendix.cidt.orig} Using 2010-era power numbers}
    \end{subfigure}
    \begin{subfigure}{0.9\columnwidth}
        \includegraphics[width=\columnwidth]{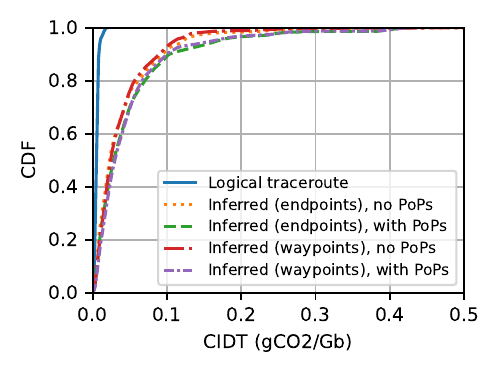}
        \caption{\label{fig:routes.appendix.cidt.new} Using 2020-era improved power numbers}
    \end{subfigure}
    \caption{\label{fig:routes.appendix.cidt.old_vs_new}Comparison of CIDT using old and new power numbers (\autoref{tab:appendix.network_power})}
\end{figure}

We show the resulting impact on CIDTs in \autoref{fig:routes.appendix.cidt.old_vs_new}.

\begin{table}[h]
\centering
\begin{tabular}{l|cc}
\hline
\multirow{2}{*}{Device} & \multicolumn{2}{c}{\begin{tabular}[c]{@{}c@{}}Energy intensity\\ (IE in W/Gbps)\end{tabular}} \\ \cline{2-3} 
                        & 2012                                               & 2024                                              \\ \hline
Core router             & 10                                                 & 2                                                 \\
WDM switch (OXC)        & 0.05                                               & 0.05                                              \\
Transceivers            & 1.5                                                & 0.09                                              \\
Amplifier               & 0.03                                               & 0.03                                              \\
Regenerator             & 3                                                  & 3                                                 \\ \hline
\end{tabular}
\caption{\label{tab:appendix.network_power}Comparison of network device energy intensity of 2012 data~\cite{van2012power} and current devices}
\end{table}

\section{Route dataset}
\label{sec:appendix.route-dataset}

We plan to open source the full result of our route dataset, but due to the large volume of data and anonymity requirement, here we show the schema and a preview of the dataset.

Between each pair routes, defined as a 4-tuple (\texttt{src\_cloud}, \texttt{src\_region}, \texttt{dst\_cloud}, \texttt{dst\_region}), and for each route accounting method (\texttt{source}), we capture:
\begin{itemize}
    \item \texttt{hop\_count}: $\#$ of hops/routers
    \item \texttt{distance\_km}: total distance in km (great circle distance for \texttt{traceroute} method)
    \item \texttt{routers\_latlon}: delimited list of routers' coordinates, in (lat, lon) format.
    \item \texttt{fiber\_wkt\_paths$^*$}: standard WKT path string, one segment per hop. Note that WKT uses (lon, lat) coordinates.
    \item \texttt{fiber\_types$^*$}: delimited list of the fiber types for each segment of \texttt{fiber\_wkt\_paths}; either \texttt{land} or \texttt{submarine}.
\end{itemize}
Note: items marked with $^*$ are only available for inferred path methods, as they come from the physical fiber map.

Preview of the dataset: 
\begin{itemize}
    \item \texttt{src\_cloud}: aws
	\item \texttt{src\_region}: us-east-1
	\item \texttt{dst\_cloud}: aws
	\item \texttt{dst\_region}: eu-north-1
	\item \texttt{hop\_count}: 9
	\item \texttt{distance\_km}: 7889.28
	\item \texttt{routers\_latlon}: (39.0469, -77.4903) | ...
	\item \texttt{fiber\_wkt\_paths}: MULTILINESTRING \\ ((-77.4903 39.0469, -77.01376 38.89731), ...)
	\item \texttt{fiber\_types}: land | ... | submarine | ... | land
	\item \texttt{source}: Inferred (waypoints), with PoPs
\end{itemize}

\section{Combined compute + transfer carbon optimization}
\label{sec:appendix.optimization-algorithm}

Because we want to optimize for the total amount of carbon emissions from both compute and transfers, we need to simultaneously optimize for three time variables instead of one: namely when to start the input transfer, when to start the compute, and when to start the output transfer.

Although it is easy to use a simple sliding window algorithm when not considering the two data transfers, it is not a straightforward task to optimize for three time variables in an efficient manner. An naive algorithm will have to scan for all possible combinations of three timing variables, which runs in $O(n^3)$ time. This does not scale with our real-time optimization requirement across multiple regions. Thus we developed a DP-based linear time algorithm by pre-computing the individual carbon emissions at any given starting time, for all three components, and then linearly scan through possible start times of the middle component, i.e. start of the compute job. Mathematically, we describe this linear time algorithm in \autoref{alg:carbon-optimization}.

\begin{algorithm*}
    \caption{An $O(n)$ algorithm for optimizing compute + transfer carbon emissions}
    \label{alg:carbon-optimization}
    \begin{algorithmic}[1]

    \Procedure{CalculateIntegral}{$f, D, T_{min}, T_{max}$}
        \LeftComment{Calculate the integral of step function f over a fixed-duration moving window, to produce the total carbon emission.}
        \State $f^I(t) \gets \int_{t}^{t + D} f(t), \forall t \in [T_{min}, T_{max}]$ \Comment{Integrate $f$ over duration $D$}
        \State \textbf{return} $f^I(t)$
    \EndProcedure

    \Procedure{GetOptimalPoints}{$f^I, T_{min}, T_{max}, reverse$}
        \LeftComment{Get the series of times at turning points of the curve $f^I$ with non-increasing values.}
        \State $OP \gets \{\}$
        \State $last\_value \gets infinity$
        \State \IfThenElse
            {$reverse$}
            {$interval \gets [-T_{max}, -T_{min}]$}
            {$interval \gets [T_{min}, T_{max}]$}

        \For{$t^{\prime} \in interval$}
            \State $t \gets \lvert t^{\prime} \rvert$
            \If{$f^{I\prime}(t - \delta) = f^{I\prime}(t + \delta)$} \Comment{Not a turning point, ignoring}
                \State \textbf{continue}
            \EndIf
            \If{$f^I(t) \leq last\_value$}
                \State $last\_value \gets f^I(t)$
                \State $OP \gets OP \vert \{ t \}$
            \EndIf
        \EndFor
        \State \textbf{return} $OP$
    \EndProcedure

    \Require $f_1, f_2, f_3: \mathbb{T} \rightarrow \mathbb{R^+}$: Step functions of carbon emission rates of input transfer, compute and output transfer
    \Require $T_0$: The initial start time
    \Require $T_4$: The maximum allowed end time
    \Require $D_1, D_2, D_3$: The duration of input transfer, compute and output transfer
    
    \Ensure $t_1, t_2, t_3$: Start times of input transfer, compute and output data transfer

    \textbf{Constraint}: $T_0 \leq t_1 < t_1 + D_1 \leq t_2 < t_2 + D_2 \leq t_3 < t_3 + D_3 \leq T_4$

    \textbf{Objective}: $argmin_{t_1, t_2, t_3} \int_{t_1}^{t_1 + D_1} f_1(t) + \int_{t_2}^{t_2 + D_2} f_2(t) + \int_{t_3}^{t_3 + D_3} f_3(t)$

    \For{$i \in \{1, 2, 3\}$}
        \State $f_i^I(t) \gets $\textsc{CalculateIntegral}($f_i, D_i, T_0 + \sum_{i^\prime = 0}^{i - 1} D_{i^\prime}, T_4 - \sum_{i\prime = i}^{3} D_{i^\prime}$)
        \If{$i = 1$}
            \State $OP_i \gets $\textsc{GetOptimalPoints}($f_i^I, T_0 + \sum_{i^\prime = 0}^{i - 1} D_{i^\prime}, T_4 - \sum_{i\prime = i}^{3} D_{i^\prime}, $\texttt{false})
        \ElsIf{$i = 3$}
            \State $OP_i \gets $\textsc{GetOptimalPoints}($f_i^I, T_0 + \sum_{i^\prime = 0}^{i - 1} D_{i^\prime}, T_4 - \sum_{i\prime = i}^{3} D_{i^\prime}, $\texttt{true})
        \EndIf
    \EndFor

    \State $min\_integral_{total} \gets $infinity
    \State $T_{optimal} \gets \{NaN, NaN, NaN\}$

    \For{$t_2 \in [T_0 + D_1, T_4 - D_2 - D_3]$} \\
        \Comment{Calculate minimum $integral_1$}
        \State $t_1\_max \gets t_2 - D_1$
        \State $op_1 \gets \max(\{op \in OP_1 | op \le t_1\_max\})$
        \State $optimal\_t_1 \gets argmin_{t \in \{op_1, t_1\_max\}} f_1^I(t)$
        \State $min\_integral_1 \gets f_1^I(optimal\_t_1)$

        \Comment{Calculate minimum $integral_3$}
        \State $t_3\_min \gets t_2 + D_2$
        \State $op_3 \gets \min(\{op \in OP_3 | op \ge t_3\_min\})$
        \State $optimal\_t_3 \gets argmin_{t \in \{op_3, t_3\_min\}} f_3^I(t)$
        \State $min\_integral_3 \gets f_3^I(optimal\_t_3)$

        \Comment{Compare total integral}
        \State $integral_{total} \gets min\_integral_1 + f_2^I(t_2) + min\_integral_3$
        \If{$integral_{total} < min\_integral_{total}$}
            \State $min\_integral_{total} \gets integral_{total}$
            \State $T_{optimal} \gets \{optimal\_t_1, t_2, optimal\_t_3 \}$
        \EndIf
    \EndFor

    \State \textbf{return} $T_{optimal}$

    \end{algorithmic}
\end{algorithm*}

\end{document}